\documentclass[sigconf]{acmart}

\renewcommand\footnotetextcopyrightpermission[1]{}
\AtBeginDocument{%
  }

\usepackage{amsmath,amsfonts}
\usepackage{bm}
\usepackage{nicefrac}
\usepackage{siunitx}
\usepackage{graphicx}
\usepackage{textcomp}
\usepackage{xcolor}
\usepackage{subcaption}
\usepackage{siunitx}
    
\usepackage[version=4]{mhchem}    
\usepackage{subcaption}

\newcommand{\rtx}{r_\mathrm{TX}}
\newcommand{\Vin}{V_\mathrm{in}}
\newcommand{\Vout}{V_\mathrm{out}}
\newcommand{\Sout}{S_\mathrm{out}}
\newcommand{\Sin}{S_\mathrm{in}}
\newcommand{\CRin}{C_{\mathrm{in}}^{\scriptsize\mathrm{R}}}

\newcommand{\CSin}{C_{\mathrm{in}}^{\scriptsize\mathrm{S}}}
\newcommand{\CSout}{C_{\mathrm{out}}^{\scriptsize\mathrm{S}}}
\newcommand{\CRoutN}{C_{\mathrm{out},0}^{\scriptsize\mathrm{R}}}
\newcommand{\CMRin}{C_{\mathrm{in}}^{\scriptsize\mathrm{MR}}}

\newcommand{\CAin}{C_{\mathrm{in}}^{\scriptsize\mathrm{A}}}
\newcommand{\CBin}{C_{\mathrm{in}}^{\scriptsize\mathrm{B}}}
\newcommand{\CAout}{C_{\mathrm{out}}^{\scriptsize\mathrm{A}}}
\newcommand{\CAoutN}{C_{\mathrm{out},0}^{\scriptsize\mathrm{A}}}
\newcommand{\CBout}{C_{\mathrm{out}}^{\scriptsize\mathrm{B}}}

\newcommand{\NBout}{N_{\mathrm{out}}^{\scriptsize\mathrm{B}}}
\newcommand{\NAin}{N_{\mathrm{in}}^{\scriptsize\mathrm{A}}}
\newcommand{\NAout}{N_{\mathrm{out}}^{\scriptsize\mathrm{A}}}
\newcommand{\NBin}{N_{\mathrm{in}}^{\scriptsize\mathrm{B}}}
\newcommand{\NBrx}{N_{\mathrm{RX}}^{\scriptsize\mathrm{B}}}
\newcommand{\Na}{N_{\mathrm{a}}}
\newcommand{\NRin}{N_{\mathrm{in}}^{\scriptsize\mathrm{R}}}
\newcommand{\NSin}{N_{\mathrm{in}}^{\scriptsize\mathrm{S}}}
\newcommand{\NSout}{N_{\mathrm{out}}^{\scriptsize\mathrm{S}}}
\newcommand{\NSrx}{N_{\mathrm{RX}}^{\scriptsize\mathrm{S}}}

\newcommand{\lAin}{\lambda_{\mathrm{in}}^{\scriptsize\mathrm{A}}}
\newcommand{\lBin}{\lambda_{\mathrm{in}}^{\scriptsize\mathrm{B}}}
\newcommand{\lBout}{\lambda_{\mathrm{out}}^{\scriptsize\mathrm{B}}}

\newcommand{\lRin}{\lambda_{\mathrm{in}}^{\scriptsize\mathrm{R}}}
\newcommand{\lSin}{\lambda_{\mathrm{in}}^{\scriptsize\mathrm{S}}}
\newcommand{\lSout}{\lambda_{\mathrm{out}}^{\scriptsize\mathrm{S}}}

\newcommand{\kab}{k_\mathrm{AB}}

\usepackage{acronym}
\acrodef{DMC}{Diffusive Molecular Communication}
\acrodef{Tx}{Transmitter}
\acrodef{Rx}{Receiver}
\acrodef{NP}{Nanoparticle}
\acrodef{ODE}{Ordinary Differential Equation}
\acrodef{IIT}{Impulse Invariant Transformation}
\acrodef{MR}{Mandelate Racemase}
\acrodef{(R)man}{(R)-Mandelate}
\acrodef{(S)man}{(S)-Mandelate}
\acrodef{PBS}{Particle Based Simulation}
\acrodef{MC}{Molecular Communication}

\begin{document}

\title{Controlled Signaling and Transmitter Replenishment for MC with Functionalized Nanoparticles}

\author{Maximilian Sch\"afer\textsuperscript{1}, Lukas Brand\textsuperscript{1}, Sebastian Lotter\textsuperscript{1}, Atakan B\"uy\"ukoglu \textsuperscript{1}, Franz Enzenhofer\textsuperscript{2}, Werner Haselmayr\textsuperscript{2}, Kathrin Castiglione\textsuperscript{1}, Dietmar Appelhans\textsuperscript{3}, Robert Schober\textsuperscript{1}}
\affiliation{%
  \institution{\textsuperscript{1}Friedrich-Alexander Universit\"{a}t Erlangen-N\"urnberg (FAU), Erlangen, Germany}
  \country{}
}
\affiliation{%
  \institution{\textsuperscript{2}Johannes Kepler University Linz (JKU),
	Linz, Austria}
\country{}
}
\affiliation{%
  \institution{\textsuperscript{3}Leibnitz-Institut f\"ur Polymerforschung Dresden e.V., Dresden, Germany}
\country{}
}

\renewcommand{\shortauthors}{Sch\"{a}fer et al.}

\begin{abstract}
In this paper, we propose novel \ac{Tx} models for \ac{MC} systems based on functionalized \acp{NP}. 
Current \ac{Tx} models often rely on simplifying assumptions for the molecule release and replenishment mechanisms. 
In contrast, we propose a \ac{Tx} model where the signaling molecule release is controlled by a switchable membrane driven by an external trigger.
Moreover, we propose a reloading mechanism, where signaling molecules are harvested based on an enzymatic reaction.
Hence, no repeated injection of signaling molecules is required.
For the proposed \ac{Tx} model, we develop a general mathematical description in terms of a discrete-time transfer function model. 
Furthermore, we investigate two realizations of the proposed \ac{Tx} model, i.e., an idealized \ac{Tx} relying on simplifying assumptions, and a realistic \ac{Tx} employing practical components for the reloading and release mechanisms.
Finally, we numerically evaluate the proposed model and compare our results to stochastic \acp{PBS}.  
\end{abstract}


\maketitle
\pagestyle{plain}

\vspace*{-0.3cm}
\section{Introduction}
\label{sec:intro}
\vspace*{-0.1cm}
\acf{MC} is a bio-inspired communication paradigm employing molecules for information exchange. 
\ac{MC} enables synthetic communication at the nanoscale, which paves the way for transformative applications in nanomedicine and health monitoring \cite{Chude_Okonkwo_19,Akyildiz_15}. 
For the design and optimization of \ac{MC} systems it is crucial to develop models for its components, i.e., the \acf{Tx}, the channel, and the \ac{Rx}. 
Various component models have been proposed, which can be roughly classified into point and volume~\acp{Tx}, diffusive, advective, and degradative channels, and active and passive \acp{Rx}~\cite{jamali:ieee:2019}. 
However, most existing models do not properly reflect the characteristics of practical components due to various simplifications. 

Specifically, the development of practical \ac{Tx} models is crucial, since the majority of the works in the \ac{MC} literature assume a point \ac{Tx}, requiring instantaneous signaling molecule production and release~\cite{jamali:ieee:2019}. 
However, realistic \ac{Tx} models have to consider the following aspects: 
(i) the signal molecule production and propagation inside the \ac{Tx}, and 
(ii) mechanisms for releasing signal molecules. 
A first step towards practical \ac{Tx} models was made in~\cite{noel:16}, which studies a virtual spherical \ac{Tx} with uniformly distributed molecules inside and a fully reflective spherical \ac{Tx} with uniformly distributed molecules on its surface, respectively. 
However, both \ac{Tx} models assume an instantaneous molecule production and release.
The works in~\cite{schaefer:icc:2020,Huang_22,Arjmandi_16,Schaefer_21} focus on the release mechanism and the propagation inside the \ac{Tx}. 
In~\cite{Arjmandi_16}, a spherical \ac{Tx} with ion-channels in its membrane is investigated, where the release of the signal molecules is controlled by the opening and closing of the ion-channels through a voltage or ligands. 
In~\cite{schaefer:icc:2020}, the molecule release from a spherical \ac{Tx} is controlled by a spatially and temporally adjustable semi-permeable membrane. 
A membrane-fusion based \ac{Tx} is considered in~\cite{Huang_22}, which encapsulates the signaling molecules in vesicles. 
The molecules are released when the vesicles fuse with the \ac{Tx} membrane. 
While the aforementioned models can be considered as reservoir-based \ac{Tx} models that control the release by membrane functionalization, in \cite{Schaefer_21} polymer matrices are considered as \acp{Tx}, which are already employed as practical drug carriers in medical applications. 
%
However, none of the works mentioned above take the production of signaling molecules into account. 
Besides the generation of molecules (e.g., via chemical reaction networks), molecule harvesting is a promising approach to generate signaling molecules. 
Recently, in \cite{Ahmadzadeh2022} a molecular harvesting \ac{Tx} is proposed, where molecules are released by release units (e.g., ion-channels~\cite{Arjmandi_16}) and re-captured if they hit a harvesting unit on the \ac{Tx} surface. 

In this paper, we propose a novel \ac{Tx} model basing on the properties of functionalized \acfp{NP}, which are promising candidates to be used as nodes in nanonetworks for sensing and localized treatment \cite{Chude_Okonkwo_19,Akyildiz_15}.
The envisioned \ac{Tx} provides a controlled molecule release mechanism based on switching the \ac{NP} membrane between the open and closed states by an external trigger, e.g., a pH change of the surrounding environment. 
Furthermore, the proposed \ac{Tx} exhibits a reloading mechanism for repeated signaling molecule production, which enables the harvesting of signaling molecules from the molecules already present in the surrounding environment of the \ac{Tx}. 
In contrast to~\cite{Ahmadzadeh2022}, the proposed reloading mechanism is based on an enzyme reaction inside the \ac{Tx} to convert molecules surrounding the \ac{Tx} to signaling molecules.
Therefore, the proposed system can also be interpreted as a realization of media modulation~\cite{Brand2022}, since the molecules that are converted to signal molecules by the \acp{NP} are already present in the environment surrounding the \ac{Tx}. Thus an external injection of signaling molecules is not required.
The main contributions of this paper are as follows:
\vspace*{-0.4cm}
\setlength{\itemsep}{0pt}
\setlength{\parskip}{0pt}
\begin{itemize}
  \item We propose a novel \ac{Tx} design for synthetic \ac{MC} systems based on functionalized \acp{NP}. The proposed \ac{Tx} enables controlled molecule release via a switchable membrane and reloading by harvesting molecules from the environment.
  \item	We develop a model for an idealized \ac{Tx} realization based on instantaneous membrane switching and a simple first order reaction for the production of the signaling molecules. 
  \item	We develop a second, more practical \ac{Tx} model based on a non-instantaneous membrane switching mechanism employing a pH sensitive polymer and the conversion of molecules from the surrounding environment to signaling molecules by a realistic enzyme reaction.
\end{itemize}
The paper is organized as follows. 
In Section~\ref{sec:system}, we introduce the proposed \ac{Tx} model for \ac{MC} systems,  including a general mathematical description. 
The general model is specialized to an idealized realization and a practical realization of the \ac{Tx} in Sections~\ref{sec:ideal} and \ref{sec:realistic}, respectively. 
Finally, Section~\ref{sec:evaluation} presents numerical results, and our main conclusions are drawn in Section~\ref{sec:conclusion}.
\vspace*{-0.3cm}

\section{System Model}
\label{sec:system}
\vspace*{-0.1cm}
\begin{figure}[t]
	\centering
	\includegraphics[width=0.9\linewidth]{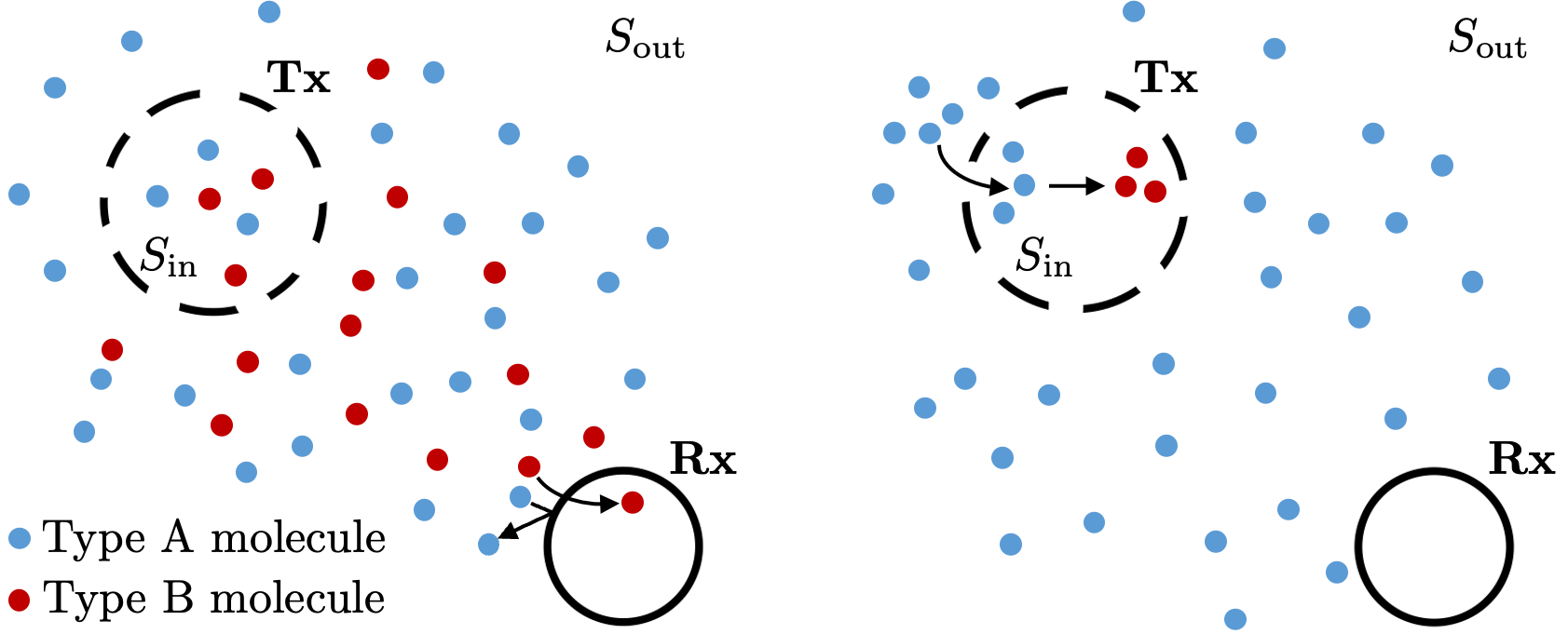}
	\vspace*{-0.4cm}
	\caption{\small Left hand side: Controlled release of signaling type B molecules (red) by controlled opening of the \ac{Tx} membrane. Right hand side: Reloading of type A molecules (blue) from the surrounding environment and conversion into signaling type B molecules.}
	\label{fig:system-design}
	\vspace*{-0.6cm}
\end{figure}

In this section, we describe the proposed \ac{Tx} design employing functionalized \acp{NP}, see Fig.~\ref{fig:system-design}. 
The envisioned \ac{Tx} comprises two main mechanisms, namely
\setlength{\itemsep}{0pt}
\setlength{\parskip}{0pt}
\vspace*{-0.1cm}
\begin{itemize}
	\item[(i)] a \textbf{controlled release mechanism}, which enables the release of signaling molecules controlled by an external stimulus (see Fig.~\ref{fig:system-design}, left hand side),
	\item[(ii)] a \textbf{reloading mechanism} enabling the production of signaling molecules inside the \ac{Tx} from molecules recruited from the surrounding environment (see Fig.~\ref{fig:system-design}, right hand side).
\end{itemize}
Both mechanisms are facilitated by the design and functionalization of the \ac{NP}. The controlled release mechanism (ii) is realized by functionalization of the \ac{NP} with a switchable membrane allowing for the control of its permeability by an external stimulus.
The reloading mechanism (i) is realized by enzymes encapsulated in the \ac{NP} allowing for the conversion of molecules recruited from the surrounding environment, denoted as type A molecules, in signaling molecules, denoted as type B molecules. 
The switching of the \ac{NP} membrane between the open and closed states is exploited for both the controlled release of type B molecules and the reloading of type A molecules, while the membrane is impermeable for enzymes in both states. 
In the following we denote the number of type A molecules in the environment, $\Sout$, and inside the \ac{NP} volume, $\Sin$, by $\NAout$ and $\NAin$, respectively (see Fig.~\ref{fig:system-design}).

To provide a tractable mathematical description of the proposed \ac{Tx}, we make the following assumptions:

\setlength{\itemsep}{-1pt}
\setlength{\parskip}{-1pt}
\begin{itemize}
	\item[A1)] The environment $S_\mathrm{out}$ with volume size $\Vout = \nicefrac{4}{3}\pi r_\mathrm{out}^3$ is much larger than $S_\mathrm{in}$ with size $\Vin = \nicefrac{4}{3}\pi r_\mathrm{in}^3$, i.e., $\Vout \gg \Vin$. 
	\item[A2)] Both volumes $\Sout$ and $\Sin$ are well mixed and $\NAout \gg \NAin$.
\end{itemize}
A1 is motivated by the fact that the typical diameter of a \ac{NP} is around hundreds of nanometer, while the diameter of a typical synthetic MC channel, e.g., a microliter reactor or a microfluidic channel, is around $\SI{1e-6}{}-\SI{1e-3}{\meter}$ \cite{Grafe2014, schaefer:laminar:2020, Schaefer_21}.
A2 is justified if the propagation of molecules due to diffusion is relatively fast compared to the absorption rate of these molecules at the boundary of the \ac{NP}. 
For $\Sout$, A2 is valid for the practical values chosen for the diffusion coefficient of the molecules, the number of molecules in $\Sout$, $\NAout$, and permeability of the \ac{NP}. 
For $\Sin$, A2 is validated by the excellent agreement of the numerical results obtained from our proposed \ac{Tx} model, relying on A2, and the results obtained from stochastic \acfp{PBS}. 

\vspace*{-0.1cm}
\subsection{Transmitter Reloading Mechanism}
\label{subsec:gr:reload}
\vspace*{-0.1cm}

The \ac{Tx} reloading is based on the diffusion of type A molecules from $\Sout$ into the \ac{NP} and their conversion into type B signaling molecules, see right hand side of Fig.~\ref{fig:system-design}. 
Based on A1 and A2, the concentration of the type A molecules inside the \ac{NP} can be modeled by the following \ac{ODE}
\begin{align}
	\partial_t \CAin(t) = -\rho(t)\left(\CAin(t) - \CAout(t)\right) - \mathcal{G}\left(\CAin(t),\, \dots\right), 
	\label{eq:genAin}
\end{align}
where $\partial_t$ denotes a derivative with respect to time $t$, and $\CAin$ and $\CAout$ denote the concentration of type A molecules in $\si{\mol\per\cubic\meter}$ in $\Sin$ and $\Sout$, respectively. 
The time-dependent permeability $\rho(t)$ in $\si{\per\second}$ models the switchable \ac{Tx} membrane\footnote{We obtained $\rho$ in $\si{\per\second}$ from practical permeability values $\hat{\rho}$, actually measured in $\si{\meter\per\second}$, by a relation to the surface area $A$ and $\Vin$ of the \ac{NP}, i.e., $\rho = \hat{\rho}A/\Vin$.}. 
The function $\mathcal{G}$ models the change of concentration $\CAin$ due to the conversion of type A molecules into type B molecules, which depends on the inner concentration $\CAin$ and the actual production mechanism. 
For example, if the production of type B molecules is realized by an enzyme reaction (see Sections~\ref{sec:ideal} and \ref{sec:realistic}), function $\mathcal{G}$ may also depend on the enzyme and cofactor concentrations inside the \ac{NP} volume $\Sin$. 

Similar to \eqref{eq:genAin}, the concentration of type B signaling molecules in $\Sin$ can be modeled by the following \ac{ODE}
\begin{align}
	\partial_t \CBin(t)= -\rho(t)\left(\CBin(t) - \CBout(t)\right) + \tilde{\mathcal{G}}\left(\CAin(t),\, \dots\right), 
	\label{eq:genBin}
\end{align}
where $\CBin$ and $\CBout$ denote the concentration of type B molecules in $\si{\mol\per\cubic\meter}$ in $\Sin$ and $\Sout$, respectively. 
Similar to function $\mathcal{G}$ in \eqref{eq:genAin}, the function $\tilde{\mathcal{G}}$ models the change of $\CBin$ due to the production of type B molecules from type A molecules.

\vspace*{-0.1cm}
\subsection{Controlled Release of Signaling Molecules}
\vspace*{-0.1cm}
The controlled release of type B molecules from the \ac{Tx} into the environment $\Sout$ depends on the \ac{NP} permeability $\rho(t)$ and on the difference between the inner and outer concentration of type B molecules which can be expressed as follows 
\begin{align}
	\partial_t \CBout(t)= \rho(t)\frac{\Vin}{\Vout}\left(\CBin(t) - \CBout(t)\right).
	\label{eq:genBout}
\end{align}
\vspace*{-0.3cm}

\subsection{Reception of Signaling Molecules}
\vspace*{-0.1cm}
As \ac{Rx} model, we adopt an absorbing spherical \ac{Rx} with radius $\rtx$ and distance $d$ from the \ac{Tx}. Moreover, the \ac{Rx} can only absorb type B molecules. 
The number of molecules $\NBrx(t)$ absorbed by the \ac{Rx} is obtained by convolving the number of molecules $\NBout(t)$, released from the \ac{Tx} surface over time, and the hitting probability $p(t)$ of an instantaneous release from the surface of a spherical \ac{Tx} previously proposed in \cite{Huang_22}, i.e., 
\vspace*{-0.15cm}
\begin{align}
	\NBrx(t) = \NBout(t)\ast p(t),
\end{align}
where $\ast$ denotes a convolution with respect to time $t$ and the number of type B molecules released by the \ac{Tx} is given by $\NBout = \CBout(t)\,\Vout\,\Na$, with Avogadro constant $\Na = 6.022\cdot 10^{23}\si{\per\mol}$. 
The analytical expression for $p(t)$ is given by \cite[Eq.~(9)]{Huang_22}. 

\vspace*{-0.1cm}

\section{Idealized Transmitter Model}
\label{sec:ideal}
\vspace*{-0.1cm}
In this section, we investigate an idealized realization of the \ac{Tx} model proposed in Section~\ref{sec:system}. 
This idealized model will serve later as a benchmark for the practical \ac{Tx} proposed in Section~\ref{sec:realistic}.
To this end, we specialize the reloading and release mechanism to obtain mathematically tractable functions. 
We assume an \textit{instantaneous switching} and a \textit{first order irreversible reaction} for type B molecule production inside the \ac{NP}.
In particular, we assume that the membrane of the \ac{NP} can be switched instantaneously between the closed and open states with permeability $\rho_\mathrm{max}$, i.e., $\rho \in \{0,\, \rho_\mathrm{max}\}$. 
We model the production of type B signaling molecules from harvested type A molecules in $\Sin$ by a simple first order reaction \cite{Noel2014}
\vspace*{-0.2cm}
\begin{align}
	&\ce{A ->[\kab] B}, &\mathcal{G}\left(\CAin(t) \right) = \partial_t \CBin(t) = \kab \CAin(t).
	\label{eq:simpReact}
\end{align}
As further idealization we assume that reaction \eqref{eq:simpReact} takes place anywhere in $\Sin$ with constant reaction rate $\kab$ in $\si{\per\second}$. 
%
With this assumption for the production of type B molecules, \eqref{eq:genAin} simplifies to
\begin{align}
 \partial_t \CAin(t) = -\rho(t)\left(\CAin(t) - \CAoutN\right) - \kab\CAin(t),
	\label{eq:simpAin}
\end{align}
where we further assume a constant type A molecule concentration in $\Sout$, i.e., $\CAout(t) = \CAoutN$. 
This assumption is justified by A1 and A2, i.e., due to the large number of molecules $\NAout$ uniformly distributed  in $\Sout$ and the volume difference between $\Sout$ and $\Sin$, the number of molecules diffusing into the \ac{NP} does not significantly change the number of type A molecules in $\Sout$.
%
Similarly, the \acp{ODE} in \eqref{eq:genBin} and \eqref{eq:genBout} simplify as follows
\begin{align}
	\partial_t \CBin(t) = -\rho(t)\,\CBin(t) + \kab\CAin(t), 
	\label{eq:simpBin}
\end{align}
and
\begin{align}
	\partial_t \CBout(t) = \rho(t)\frac{\Vin}{\Vout}\CBin(t),  
	\label{eq:simpBout}
\end{align}
where we exploited A1 and A2 to assume an irreversible transmission of type B molecules through the Tx membrane, where the influx of type B molecules from the environment into the Tx is neglected (perfect sink).
\vspace*{-0.1cm}

\subsection{Discrete-time Transfer Function Model}
\label{subsec:sr:tfm}
\vspace*{-0.1cm}
Due to the time-dependent permeability $\rho(t)$ in \eqref{eq:simpAin}--\eqref{eq:simpBout}, no closed form solution in the continuous-time domain can be derived. 
%
Therefore, we derive a discrete-time transfer function model for the system of \acp{ODE} \eqref{eq:simpAin}--\eqref{eq:simpBout} via the \ac{IIT} \cite{schaefer:laminar:2020}.
Applying the \ac{IIT} to \ac{ODE} \eqref{eq:simpAin} yields a scalar valued discrete-time state equation for the concentration of type A \mbox{molecules in $\Sin$}
\begin{align}
	\CAin[k] = \exp\left(\lAin[k] T\right)\CAin[k-1] + T \rho[k] \CAoutN,
	\label{eq:simpAin-d}
\end{align}
with discrete-time index $k$, sampling time $T$ (i.e., $t = kT)$, and time-dependent parameter $\lAin[k] = -\left(\rho[k] + \kab\right)$. 

\noindent Similarly, applying an \ac{IIT} to \eqref{eq:simpBin} and \eqref{eq:simpBout} yields
%
\begin{align}
	\CBin[k] = \exp\left(\lBin[k] T\right)\CBin[k-1] + T\kab\CAin[k], 
	\label{eq:simpBin-d}
\end{align}
and 
\begin{align}
	\CBout[k] = \CBout[k-1] + \lBout[k]\CBin[k],
	\label{eq:simpBout-d}
\end{align}
with $\lBin[k] = - \rho[k]$ and $\lBout[k] = - \rho[k]{\Vin}/{\Vout}$.
%
The discrete-time solution in \eqref{eq:simpAin-d}--\eqref{eq:simpBout-d} has several benefits. 
First, the description in the discrete-time domain allows the incorporation of the time-variance introduced by permeability $\rho(t)$ in terms of time-variant parameters $\lAin$ and $\lBin$. 
Second, the proposed model allows a simple and computationally efficient solution of the \acp{ODE} in \eqref{eq:simpAin}--\eqref{eq:simpBout}.



\section{Practical Transmitter Model}
\label{sec:realistic}
\begin{figure*}
	\centering
	\includegraphics[width=0.75\linewidth]{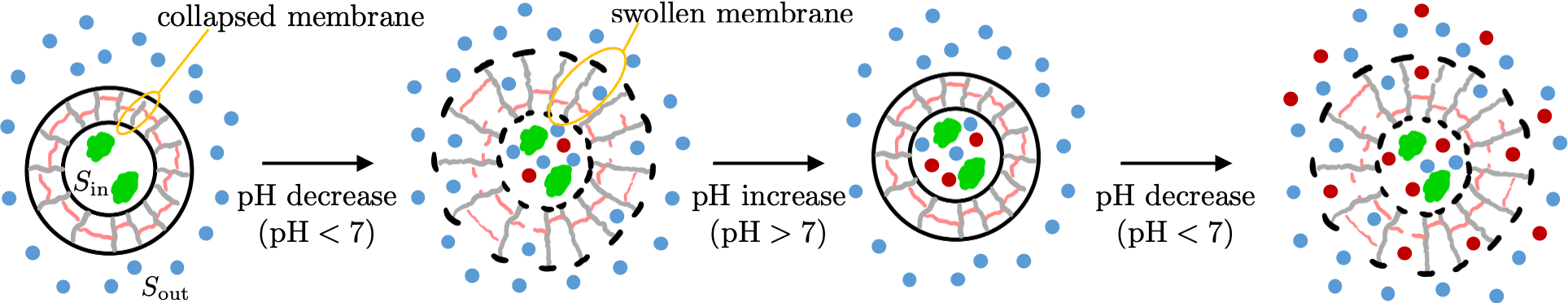}
	\vspace*{-0.5cm}
	\caption{\small Reloading and release process of the practical \ac{Tx} model. From left to right: 
	Only \acs{(R)man} molecules (blue) and enzymes (green) are present in the environment; 
	pH-decrease triggers the opening of the \ac{Tx} membrane and \acs{(R)man} molecules diffuse in $\Sin$ and are converted to \acs{(S)man} signaling molecules (red); pH-increase triggers closing of the \ac{Tx} membrane; pH-decrease triggers opening of the \ac{Tx} membrane and the produced \acs{(S)man} molecules are released while additional \acs{(R)man} molecules are harvested.}
	\label{fig:proc-overview}
	\vspace*{-0.4cm}
\end{figure*}

\vspace*{-0.1cm}
In this section, we consider a realization of the \ac{Tx} model proposed in Section~\ref{sec:system} based on practical components for the production and controlled release of signaling molecules.

\vspace*{-0.1cm}
\subsection{Membrane Switching Mechanism}
\label{subsec:real:switch}
\vspace*{-0.1cm}
The key property of the proposed \ac{Tx} for both controlled release and reloading is the ability to open and close the \ac{NP} membrane upon a controllable trigger signal.
As a practical type of \ac{NP}, we employ polymersomes, the synthetic counterpart of liposomes, with a pH-driven permeability switch, which consist of a multifunctional amphiphilic block copolymer.
In particular, it consists of the biocompatible poly(ethylene glycol) (PEG) as the hydrophilic part and a statistical mix of two components in the hydrophobic part providing the pH sensitive membrane switching functionalities, see \cite{Gaitzsch2012,Grafe2014} for details.
For high pH values, i.e., $\mbox{pH} > 7$, the membrane of the \ac{NP} collapses, i.e., the membrane is non-permeable and molecules can neither diffuse out of the \ac{NP} nor into it (permeability $\rho = 0$). 
When the pH value is low, i.e., $\mbox{pH} < 7$, the membrane swells and becomes permeable with permeability $\rho_\mathrm{max}$. 
Fig.~\ref{fig:proc-overview} shows the switching process of the \ac{NP} for multiple pH changes.

Compared to the idealized scenario, considered in Section~\ref{sec:ideal}, for practical \acp{NP}, the membrane switching is no longer instantaneous. 
In particular, upon the external pH trigger it takes approximately $30$--$100 \si{\second}$ until the membrane permeability reaches $\rho_\mathrm{max}$, see \cite[Fig.~3]{Gaitzsch2012}. 
A similar time delay applies for the closing process upon a pH increase. 
We note that the exact temporal evolution of the \ac{NP} permeability $\rho(t)$ is not known. 
Therefore, the change of $\rho(t)$ from $\rho = 0$ (closed) to $\rho = \rho_\mathrm{max}$ (open) and vice versa is modeled by a linear increase and a linear decrease, respectively. 
Also the maximum permeability in the open state is not given in a closed form, but we estimated $\rho_\mathrm{max}$ from the measurements provided in \cite{Gaitzsch2012support} employing the methods from \cite{Battaglia2006}. 
\vspace*{-0.1cm}

\subsection{Production of Signaling Molecules}
\vspace*{-0.1cm}
In the idealized scenario discussed in Section~\ref{sec:ideal}, we employed a first order reaction for signaling molecule production. 
In this section, we investigate a realistic mechanism for the production of signaling molecules by an enzyme reaction. 

We realize the production of signaling molecules by the encapsulation of \textit{\ac{MR}} into the \ac{NP}. 
The enzyme \ac{MR} performs a reversible one-substrate-reaction of \textit{$(R)$-Mandelate} (type A molecules) to \textit{$(S)$-Mandelate} (type B signaling molecules), and the reaction equations are as follows \cite{Narmandakh2010, Maurice2002}
\begin{align}
	\ce{E + (R)man <=>[$k_1$][$k_{-1}$] E*(R)man <=>[$k_2$][$k_{-2}$] E*(S)man 
	<=>[$k_3$][$k_{-3}$]E + (S)man},
	\label{eq:mandelate-racemase}
\end{align}
where E denotes the enzyme \ac{MR}, and E$\cdot$(R)man and E$\cdot$(S)man denote intermediate complexes of \ac{MR} and \acf{(R)man} and \acf{(S)man}, respectively. 
The corresponding reaction rates are denoted by $k_1, \,k_2, \,k_3$ and $k_{-1},\, k_{-2}, k_{-3}$, respectively. 
The production of \ac{(S)man} molecules from \ac{(R)man} molecules harvested from the surrounding environment is illustrated in Fig.~\ref{fig:proc-overview}. 

\ac{MR} exhibits several benefits making it a suitable enzyme for the practical realization of the proposed \ac{Tx}.
First, \ac{MR} does not require a cofactor for the reaction to take place.
This property enables a longer operating life of the \ac{Tx} in practice, as \ac{MR} produces  signaling molecules as long as molecules can be harvested and is not bound by the number of cofactor molecules encapsulated into the \ac{NP}.
Second, \ac{MR} is inexpensive and easy to synthesize in practice, which is beneficial for building a synthetic MC testbed in future work \cite{Narmandakh2010, Maurice2002}.
However, compared to the simple first order reaction in \eqref{eq:simpReact}, reaction \eqref{eq:mandelate-racemase} is bidirectional and does not convert all \ac{(R)man} into \ac{(S)man}. 
Instead the concentration of both reaches an equilibrium, i.e., half of the harvested molecules are converted into signaling molecules (see Fig.~\ref{fig:proc-overview}) \cite{Maurice2002}. 
As reaction equation \eqref{eq:mandelate-racemase} of \ac{MR} is more complex than the simple reaction equation in \eqref{eq:simpReact} it cannot be solved in a closed form nor simplified further \cite{Maurice2002}.

We note that the enzyme chosen in the \ac{Tx} also affects the \ac{Rx} design.
For \ac{MR}, a practical absorbing \ac{Rx} may comprise \textit{(S)-Mandelate Dehydrogenase} which can oxidize \ac{(S)man} signaling molecules into \textit{benzoylformate} \cite{Lehoux1999}. 
However, the detailed investigation of a practical \ac{Rx} is left for future work. In this paper we adopt the absorbing \ac{Rx} model introduced in Section~\ref{sec:system}.

\vspace*{-0.1cm}
\subsection{Discrete-time Transfer Function Model}
\vspace*{-0.1cm}

Similar to the discrete-time model derived for the idealized \ac{Tx} model in Section~\ref{subsec:sr:tfm}, we now derive a discrete-time model for the practical realization of the proposed \ac{Tx}.  

Applying the \ac{IIT} to the equations for the reloading mechanism \eqref{eq:genAin} and \eqref{eq:genBin}, we obtain a non-linear discrete-time state equation for the concentration of \ac{(R)man} in $\Sin$ as follows
\begin{align}
	\CRin[k] &= \exp(\lRin[k] T)\,\CRin[k-1] + T\rho[k]\CRoutN - \nonumber\\&T\mathcal{R}\left(\CRin[k-1], \,\CSin[k-1], \,\CMRin[k-1], \bm{k} \right),	
	\label{eq:cr:rin}
\end{align}
with $\lRin[k] = -\rho[k]$. 
Concentrations $\CRin$, $\CRoutN$ and $\CMRin$ denote the concentration of \ac{(R)man} in $\Sin$ and $\Sout$, and the concentration of \ac{MR} in $\Sin$, respectively. 
Vector $\bm{k} = \left[k_1, k_2, k_3, k_{-1}, k_{-2}, k_{-3} \right]$ comprises all reaction rates of \ac{MR}. 
The non-linear function $\mathcal{R}$ describes the change of $\CRin$ due to the enzyme reaction \eqref{eq:mandelate-racemase} facilitated by \ac{MR}. 
Similarly, we obtain for the concentration of the \ac{(S)man} signaling molecules in $\Sin$
\begin{align}
	\CSin[k] &= \exp(\lSin[k] T)\,\CSin[k-1] + \nonumber\\&T\mathcal{S}\left(\CRin[k-1], \,\CSin[k-1], \,\CMRin[k-1], \bm{k} \right),
	\label{eq:cr:sin}	
\end{align} 
with $\lSin[k] = -\rho[k]$ and the non-linear function $\mathcal{S}$ modeling the change of $\CSin$ due to the enzyme reaction.
As previously mentioned, the reaction \eqref{eq:mandelate-racemase} cannot be solved analytically, i.e., we cannot express the number of \ac{(R)man} molecules produces over time in terms of the number of \ac{(S)man} molecules and vice versa. 
Therefore, functions $\mathcal{R}$ in \eqref{eq:cr:rin} and $\mathcal{S}$ in \eqref{eq:cr:sin} cannot be given in a closed form. 
Instead, $\mathcal{R}$ and $\mathcal{S}$ represent the numerical solution of \eqref{eq:mandelate-racemase} to calculate the change of $\CRin$ and $\CSin$ in each time step $k$ based on the previous time step. 
In general, a numerical solution for \eqref{eq:mandelate-racemase} can be obtained by the numerical methods described, e.g., in \cite{Higham2008}.
We obtain a numerical solution for \eqref{eq:mandelate-racemase} as follows: 
First, we assume a pseudo steady state and derive a reaction rate equation for the three individual bidirectional reactions in \eqref{eq:mandelate-racemase} see, e.g., \cite[Sec.~7]{Higham2008} and \cite{Segel1989}.
Second, we solve these individual reactions by an exponential approach and calculate them sequentially in each time step\footnote{Please find further information and the Python implementation used for numerical evaluation on: \url{https://www.maximilianschaefer.org/publication/nanocom22/}}.
%
%
%




The concentration $\CSout$ of \ac{(S)man} molecules in $\Sout$ adopts a similar shape as for the idealized realization in \eqref{eq:simpBout-d} as follows
\begin{align}
	\CSout[k] = \CSout[k-1] + \lSout[k]\CSin[k],
	\label{eq:cr:sout}
\end{align}
with $\lSout[k] = \rho[k]{\Vin}/{\Vout}$.

\section{Numerical Evaluation}
\label{sec:evaluation}
\vspace*{-0.1cm}
In this section, numerical results obtained for the models proposed in Sections~\ref{sec:ideal} and \ref{sec:realistic} are presented along with results obtained from stochastic \acp{PBS}. 
The default parameter values are listed in Table~\ref{tab:sim_params}. 
\vspace*{-0.15cm}
\subsection{Particle Based Simulation}
\vspace*{-0.1cm}
For \ac{PBS} we adopted the simulator design in \cite{Schaefer_21}. 
The simulator features three dimensional Brownian motion of type A and type B molecules. 
Both the first order reaction and the \ac{MR} based reaction inside the \ac{NP} are implemented using the methods in \cite{Noel2014, Andrews2004}. 
The propagation of molecules through the permeable membrane is realized by exploiting the concept for partial transmission presented in \cite[Sec.~4.5]{Andrews2010}.
Due to the large number of molecules $\NAout = 10^{16}$ in the environment, it is not feasible to track the position of all molecules. 
Therefore, the \ac{PBS} only considers the molecules entering the \ac{NP} during the simulation time by realizing the influx of molecules into the \ac{NP} by a Monte Carlo simulation. 
In particular, the number of molecules entering the \ac{NP} per time step follows a Binomial distribution with the time-dependent mean value $\rho[k]\Vin\CAoutN\Na T$.
\vspace*{-0.4cm}


\begin{table}
    \vspace*{0.07in}
    \centering
    \caption{Parameter values for simulation.}
    \vspace*{-0.15in}
    \footnotesize
    \begin{tabular}{| p{.11\linewidth} | r | p{.35\linewidth} | c|}
        \hline Parameter & Default Value & Description & Ref.\\ \hline
        $D$ & $\SI{2.6e-12}{\meter\squared\per\second}$& Diffusion coefficient & \cite{Caldero2000}\\ \hline
        $\kab$ &  $10^{-1}\si{\per\second}$ & Reaction rate $A \to B$ &\\\hline
        $k_1$ & $\SI{3.21e3}{\cubic\meter\per\mol\per\second}$& Reaction rate \ac{MR}& \cite{Maurice2002}\\\hline
        $k_{-1}$ & $\SI{3948}{\per\second}$& Reaction rate \ac{MR}& \cite{Maurice2002}\\\hline
        $k_2$ & $\SI{889}{\per\second}$& Reaction rate \ac{MR}& \cite{Maurice2002}\\\hline
        $k_{-2}$ &$\SI{631.41}{\per\second}$ & Reaction rate \ac{MR}& \cite{Maurice2002}\\\hline
        $k_3$ & $\SI{3896}{\per\second}$& Reaction rate \ac{MR}& \cite{Maurice2002}\\\hline
        $k_{-3}$ & $\SI{4.46e3}{\cubic\meter\per\mol\per\second}$& Reaction rate \ac{MR}& \cite{Maurice2002}\\\hline
        $r_\mathrm{in}$ & $\SI{80e-9}{\meter}$ & \ac{NP} radius & \cite{Gaitzsch2012}\\\hline
        $r_\mathrm{out}$ & $\SI{1e-3}{\meter}$ & Radius of surrounding volume & \\\hline
        $r_\mathrm{RX}$ & $\SI{1e-6}{\meter}$& Receiver radius &\\\hline
        $d$ & $\SI{2e-6}{\meter}$ & \ac{Tx}-\ac{Rx}-Distance &\\\hline
        $\rho_\mathrm{max}$ & $\SI{2.7e-2}{\per\second}$ & Maximum permeability & \cite{Gaitzsch2012support,Battaglia2006}\\\hline
        $\NAout$ & $\SI{1e16}{}$ & Number of molecules in the surrounding environment& \\\hline
        $T$ & $\SI{1e-4}{\second}$ & Simulation time step & \\\hline
    \end{tabular}
    \label{tab:sim_params}
    \vspace*{-0.3cm}
\end{table}

\vspace*{-0.1cm}
\subsection{Evaluation of the Idealized \ac{Tx} Realization}
\vspace*{-0.1cm}
In this section, we investigate the idealized \ac{Tx} model described in Section~\ref{sec:ideal}. 
The idealized model has several parameters to influence the reloading and release mechanism, i.e., the reaction rate $\kab$, membrane switching times $t_\mathrm{s}$ and $\rho_\mathrm{max}$. 
However, in this paper, we only vary the reaction rate $\kab$ and keep the switching times $t_\mathrm{s}$ and permeability $\rho_\mathrm{max}$ fixed. 
For comparability, the default reaction rate $\kab = 10^{-1}\si{\per\second}$ is chosen to produce approximately the same amount of signaling molecules per second as the practical \ac{Tx} model.

\vspace*{-0.1cm}
\paragraph{Reloading Mechanism} 
\begin{figure}[t]
\centering
\includegraphics[width=\linewidth]{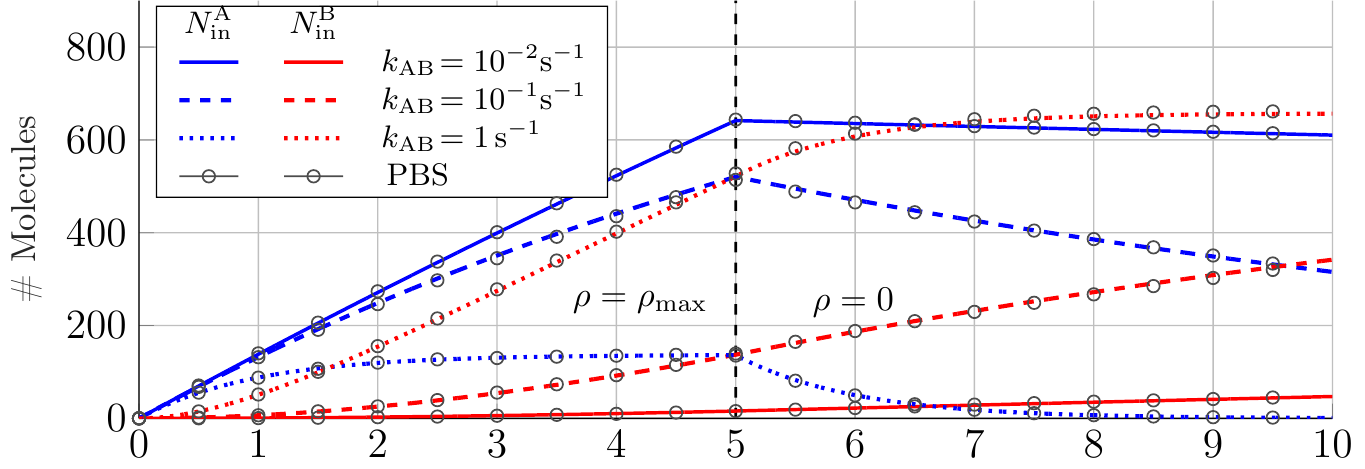}\\[-0.3em]
\includegraphics[width=\linewidth]{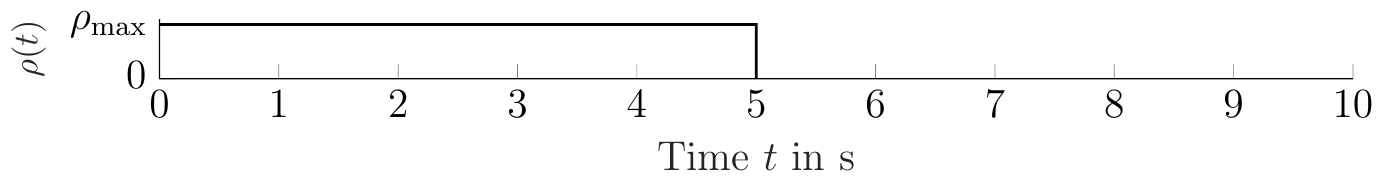}
\vspace*{-0.75cm}
\caption{\small Top: Number of type A molecules $\NAin$ and type B molecules $\NBin$ inside the \ac{Tx} over time for different reaction rates $\kab$. Bottom: Membrane switching pattern.}
\label{fig:sr:production}
\vspace*{-0.5cm}
\end{figure}
The top of Fig.~\ref{fig:sr:production} shows the amount of type A (blue curves) and type B molecules (red curves) inside the \ac{Tx} for different reaction rates $\kab$ for the membrane switching pattern shown at the bottom with $t_\mathrm{s} \in \{0, 5\}\si{\second}$.   
We observe that the number of type A molecules $\NAin$ increases during the open phase of the membrane and molecules can diffuse into the \ac{Tx} from the environment. 
When the membrane is closed ($\rho = 0$), $\NAin$ decreases because no molecules enter the \ac{Tx} and the harvested type A molecules are converted into type B molecules.
%
Next, we investigate the production of type B molecules and the impact of reaction rate $\kab$. 
The red curves in Fig.~\ref{fig:sr:production} show the amount of type B molecules $\NBin$ inside the \ac{Tx} for different reaction rates $\kab$. 
We observe, that the number of type B molecules $\NBin$ produced per second increases for increasing $\kab$ because the probability that a harvested type A molecule is converted into a type B molecule increases. 
For $\kab = 10^{-2}\si{\per\second}$ (dashed line) and $\kab = 10^{-1}\si{\per\second}$ (solid lines), $\NAin$ increases faster than $\NBin$ during the open phase. 
This behavior can be explained by inspecting \eqref{eq:simpAin}: For a small reaction rate $\kab$, the influx of type A molecules (first term on the left hand side of \eqref{eq:simpAin}) is larger than the amount of type A molecules converted into type B molecules (second term on the right hand side of \eqref{eq:simpAin}). 
Next, we observe that the $\NAin$ saturates around $t = 2\si{\second}$, while $\NBin$ increases linearly for $\kab = 1\si{\per\second}$ (dotted lines). 
In this case, the production of type B molecules and the harvesting of type A molecules reaches an equilibrium, i.e., for $t > 2\si{\second}$, $\NAin$ does not increase further because all type A molecules entering the \ac{Tx} are converted into type B molecules. 
Finally, we observe that only for $\kab = \SI{1}{\per\second}$, the reaction is fast enough to convert all harvested type A molecules into type B molecules during the closed phase. 

\vspace*{-0.1cm}
\paragraph{Multiple Switching Phases and Reception}
\begin{figure}[t]
	\centering
	\includegraphics[width=\linewidth]{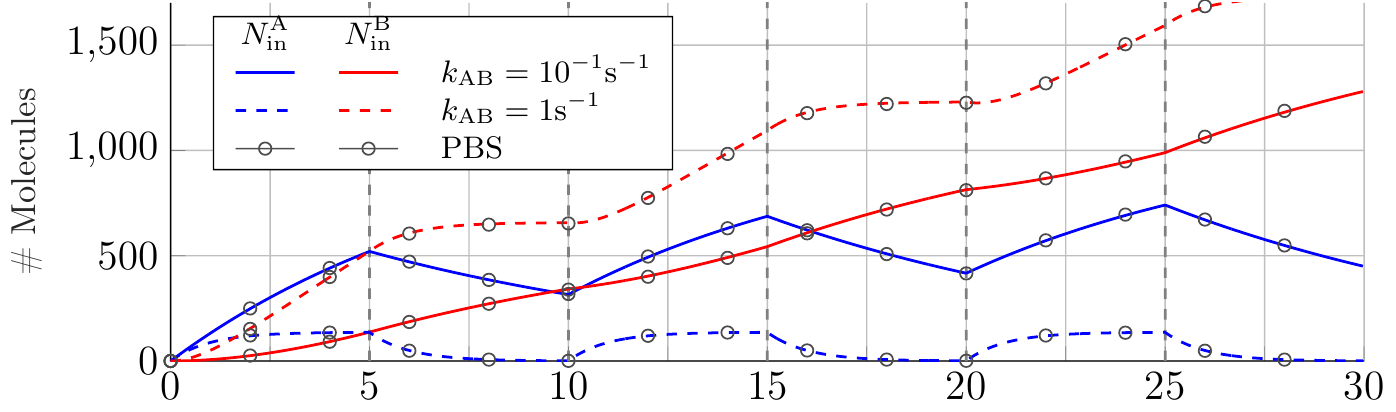}\\[-0.3em]
	\includegraphics[width=\linewidth]{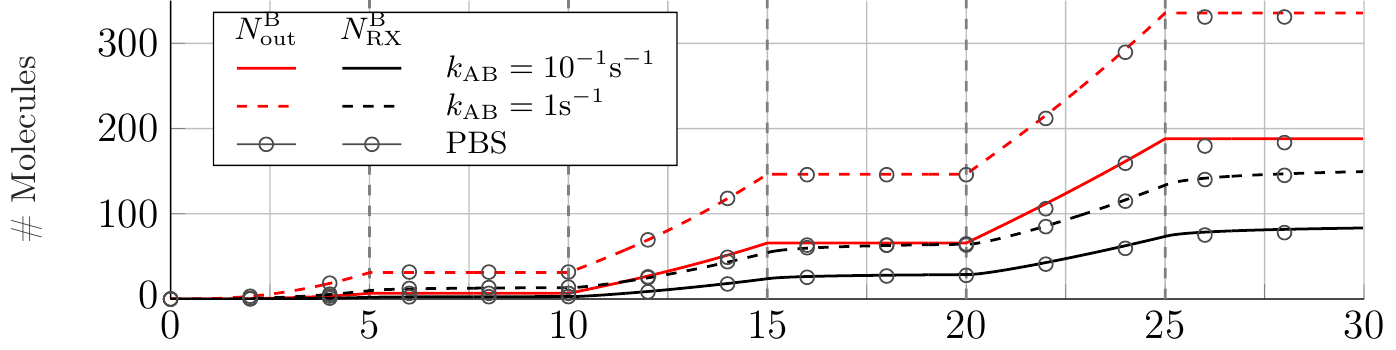}\\[-0.3em]
	\includegraphics[width=\linewidth]{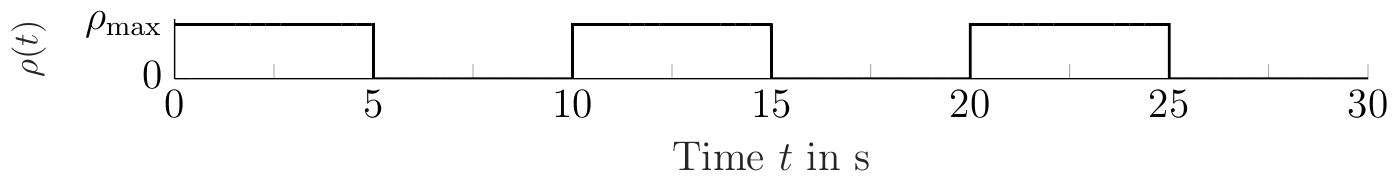}
	\vspace*{-0.7cm}
	\caption{\small Top: Current number of type A molecules $\NAin$ and type B molecules $\NBin$ inside the \ac{Tx}. Center: Accumulated number of released molecules $\NBout$ and number of received molecules $\NBrx$ at the \ac{Rx}. Bottom: Membrane switching pattern.}	
	\label{fig:sr:multiple}
	\vspace*{-0.6cm}
\end{figure}
Now, we consider multiple switching intervals to investigate repeated reloading phases and molecule releases.
The top plot of Fig.~\ref{fig:sr:multiple} shows the reloading of type A molecules $\NAin$ and the production of signaling molecules $\NBin$ for two different reaction rates $\kab$. 
The center plot shows the accumulated number of signaling molecules released from the \ac{Tx}, $\NBout$, and the accumulated number of signaling molecules absorbed by the \ac{Rx}, $\NBrx$. 
The bottom plot of Fig.~\ref{fig:sr:multiple} shows the membrane switching pattern for $t_\mathrm{s}\in\{0, 5, 10, 15, 20, 25\}\si{\second}$.

First, we observe that the behavior of the individual reloading processes is similar to that observed in Fig.~\ref{fig:sr:production}. 
Next, we observe from the top plot of Fig.~\ref{fig:sr:multiple} that the number of type B  molecules $\NBin$ inside the \ac{Tx} increases over the switching intervals. 
This effect has two reasons. First, due to the small permeability $\rho_\mathrm{max}$ the amount of released type B molecules $\NBout$ is smaller than the amount of type B molecules produced (see \eqref{eq:simpBin}). 
Second, the maximum capacity of the \ac{Tx}, which is around $N_\mathrm{max} = \CAoutN \Vin \Na \approx 5\cdot 10^{3}$, is not reached for the considered switching pattern. 
Each opening of the \ac{NP} membrane not only starts a reloading process but also triggers a controlled release of signaling molecules from the \ac{Tx}. 
To investigate the signaling properties, the center plot of Fig.~\ref{fig:sr:multiple} shows the amount of type B molecules released from the \ac{Tx} $\NBout$ (red curves) and received by the \ac{Rx} $\NBrx$ (black curves). 
First, we observe that the individual releases of type B molecules from the \ac{Tx} are clearly distinguishable (red curves). 
This clear differentiation between individual releases is due to the instantaneous switching of the \ac{Tx} membrane which immediately starts and stops the transport of molecules when the \ac{Tx} membrane opens and closes. 
This also leads to a clearly distinguishable reception of the individual releases at the \ac{Rx} (black curves). 
Moreover, we observe from Fig.~\ref{fig:sr:multiple} that the amount of released $\NBout$ and received molecules $\NBrx$  increases for an increasing reaction rate $\kab$. 
This effect is directly related to the rate dependent production of type B signaling molecules inside the \ac{Tx} as discussed in the previous section, cf. Fig.~\ref{fig:sr:production}. 
Finally, we note that all results obtained with our model and the results from \ac{PBS} match very well.
\vspace*{-0.2cm}
%
%

\vspace*{-0.1cm}
\subsection{Evaluation of the Practical \ac{Tx} Realization}
\vspace*{-0.1cm}
\begin{figure}[t]
	\centering
	\includegraphics[width=\linewidth]{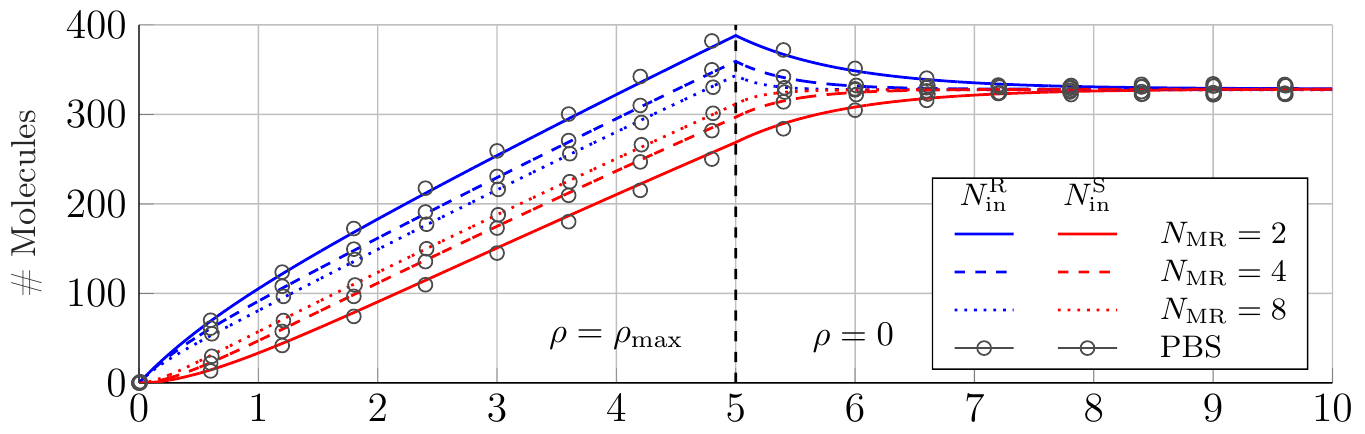}\\[-0.3em]
	\includegraphics[width=\linewidth]{img/plot1-complex-control.pdf}
	\vspace*{-0.7cm}
	\caption{\small Number of \ac{(R)man} molecules $\NRin$ and \ac{(S)man} molecules $\NSin$ inside the \ac{Tx} for different number of encapsulated enzyme $N_\mathrm{MR}$.}
	\label{fig:cr:production}
	\vspace*{-0.6cm}
\end{figure}

In this section, we investigate the practical realization of the proposed \ac{Tx} model described in Section~\ref{sec:realistic}.
The practical model has two parameters to influence the reloading and release mechanism, i.e., the membrane switching times $t_\mathrm{s}$ and the number of encapsulated enzymes $N_\mathrm{MR}$. 
The permeability $\rho_\mathrm{max}$ in the open state cannot be influenced as it depends on the practical components, e.g., the polymers, used for \ac{NP} fabrication, see Section~\ref{subsec:real:switch} and \cite{Grafe2014, Gaitzsch2012}. 
For the considered \ac{NP}, the membrane permeability is controlled by the pH-value of $\Sout$, cf. Section~\ref{subsec:real:switch}. 
In the following, we assume that the pH in $\Sout$ can be controlled perfectly, and therefore, we express different switching patterns in terms of the permeability $\rho(t)$.
\vspace*{-0.15cm}
\paragraph{Reloading Mechanism} 
First, we investigate the signaling molecule production facilitated by \ac{MR}, and therefore, we employ the instantaneous switching pattern from Fig.~\ref{fig:sr:production}.
Fig.~\ref{fig:cr:production} shows the amount of \ac{(R)man} (blue curves) and \ac{(S)man} molecules (red curves) inside the \ac{Tx} for different number of encapsulated \ac{MR} enzymes. 
First, we observe that the number of \ac{(R)man} $\NRin$ (blue curves) and signaling \ac{(S)man} molecules $\NSin$ (red curves) inside the \ac{Tx} reach an equilibrium. 
This effect is different compared to the properties of the previously discussed idealized \ac{Tx} model. 
While the first order reaction \eqref{eq:simpAin} converts all molecules harvested from the environment into signaling molecules (see Fig.~\ref{fig:sr:production}), the reaction facilitated by \ac{MR} in \eqref{eq:mandelate-racemase} pursues an equilibrium between \ac{(R)man} and \ac{(S)man} molecules \cite{Maurice2002}.
Now, we investigate the impact of the number of enzymes encapsulated into the \ac{NP}. 
We observe that for an increasing number of enzymes the equilibrium is reached faster, i.e., for $N_\mathrm{MR} = 2$ the equilibrium is reached $\SI{4}{\second}$ after closing (solid lines), while it is reached $\SI{1}{\second}$ after closing for $N_\mathrm{MR} = 8$ (dotted lines).
However, the number of enzymes does not influence the number of produced \ac{(S)man} molecules. 
Therefore, after \ac{Tx} reloading,  half of the harvested molecules are available as signaling molecules for transmission independent \mbox{of $N_\mathrm{MR}$.}

\begin{figure}[t]
	\centering
	\includegraphics[width=\linewidth]{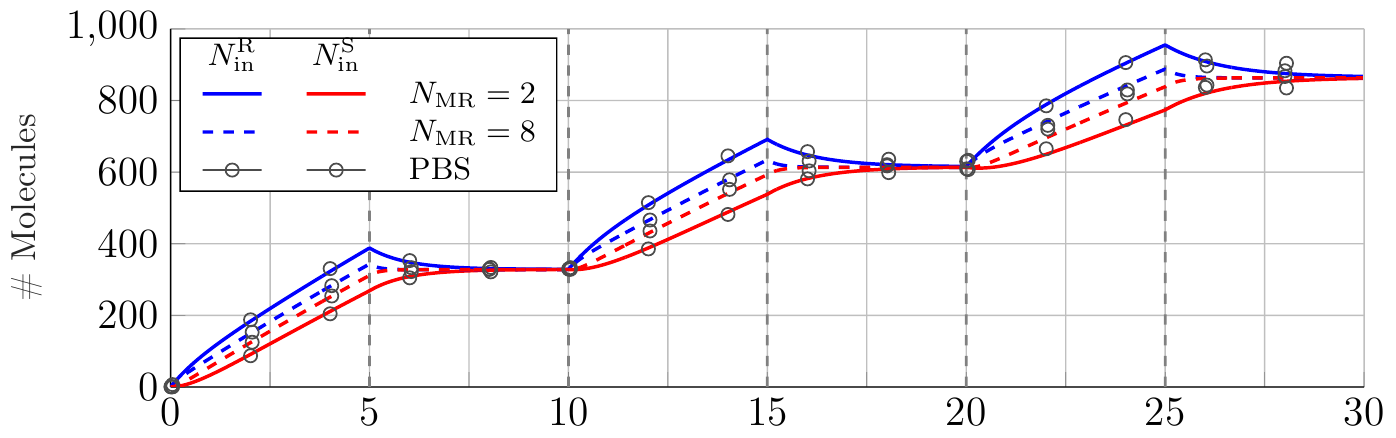} \\[-0.3em]
	\includegraphics[width=\linewidth]{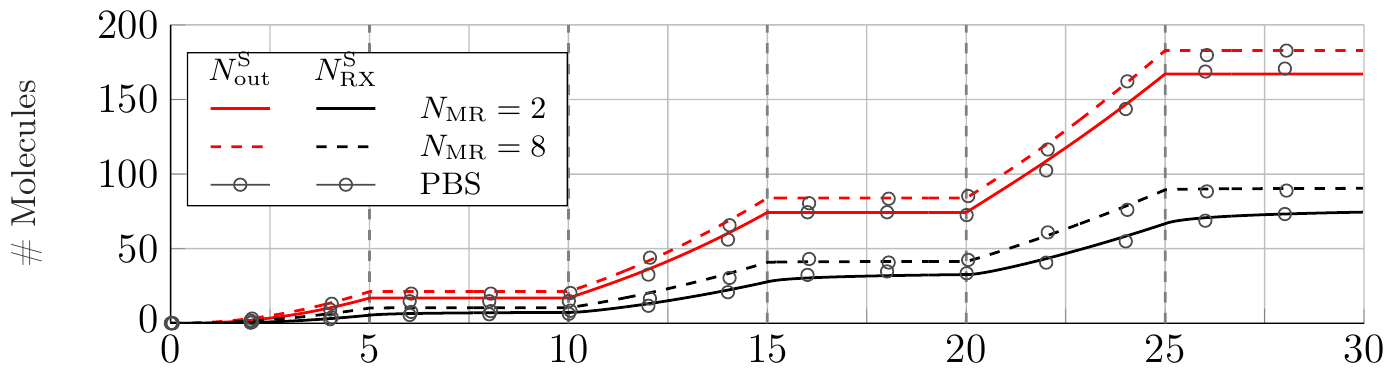}\\[-0.3em]
	\includegraphics[width=\linewidth]{img/plot2-simple-control.pdf}
	\vspace*{-0.7cm}
	\caption{\small Top: Number of \ac{(R)man} $\NRin$ and \ac{(S)man} molecules $\NSin$ inside the \ac{Tx}. Center: Number of molecules released by the \ac{Tx} $\NSout$ and received at the \ac{Rx} $\NSrx$. Bottom: Membrane switching pattern.}
	\label{fig:cr:multiple}
	\vspace*{-0.7cm}
\end{figure}

\vspace*{-0.2cm}
\paragraph{Multiple Switching Phases and Reception}
Similar to the idealized realization, we now consider multiple switching intervals. 
The top plot of Fig.~\ref{fig:cr:multiple} shows the number of \ac{(R)man} and \ac{(S)man} molecules inside the \ac{Tx}. 
The center plot shows the accumulated number of \ac{(S)man} molecules released from the \ac{Tx}, $\NSout$, and the accumulated number of \ac{(S)man} molecules absorbed at the \ac{Rx}, $\NSrx$. 
In order to make these results comparable to those from Fig.~\ref{fig:sr:multiple}, we applied the same instantaneous switching pattern (see bottom plot of Fig.~\ref{fig:cr:multiple}). 
First, we observe that behavior of each individual switching process is similar to that observed in Fig.~\ref{fig:cr:production}. 
In particular, whenever the \ac{Tx} membrane is closed an equilibrium is reached between \ac{(R)man} and \ac{(S)man}. 
Next, we observe that the individual releases of signaling \ac{(S)man} molecules $\NSout$ are clearly distinguishable. 
Similar to the idealized \ac{Tx} model, this effect is due to the instantaneous membrane switching pattern. 
Moreover, we observe that the amount of molecules released from the \ac{Tx} $\NSout$ and received by the \ac{Rx} $\NSrx$ in each opening phase of the \ac{Tx} membrane is approximately half as large as the amount released by the idealized \ac{Tx}, see Fig.~\ref{fig:sr:multiple}. 
This observation is consistent with the results shown in Fig.~\ref{fig:cr:production}. 
In particular, \ac{MR} does not convert all harvested molecules into signaling molecules but pursues an equilibrium between \ac{(R)man} and \ac{(S)man} inside the \ac{Tx}. 
Therefore, only half of the harvested molecules are available as signaling molecules. 
Finally, we observe that the number of encapsulated enzyme $N_\mathrm{MR}$ only has minor influence on the number of released molecules, but it controls the time until the equilibrium is reached inside the \ac{Tx}, see also Fig.~\ref{fig:cr:production}.
This is plausible, because increasing the number of enzymes increases the probability that a harvested molecule reacts, but does not influence the reaction rates nor the equilibrium, see \eqref{eq:mandelate-racemase}. 
However, Fig.~\ref{fig:cr:multiple} shows that $N_\mathrm{MR}$ is an important design parameter for the proposed \ac{Tx}. In particular, increasing $N_\mathrm{MR}$ decreases the time until the maximum number of signaling molecules is produced, and therefore, the duration of the closed phase between two transmission can be reduced. 
Finally, we note that the results obtained from the proposed model and the results obtained from \ac{PBS} match very well. 


%

\vspace*{-0.1cm}
\paragraph{Influence of Non-Instantaneous Switching}
\begin{figure}[t]
	\centering
	\includegraphics[width=\linewidth]{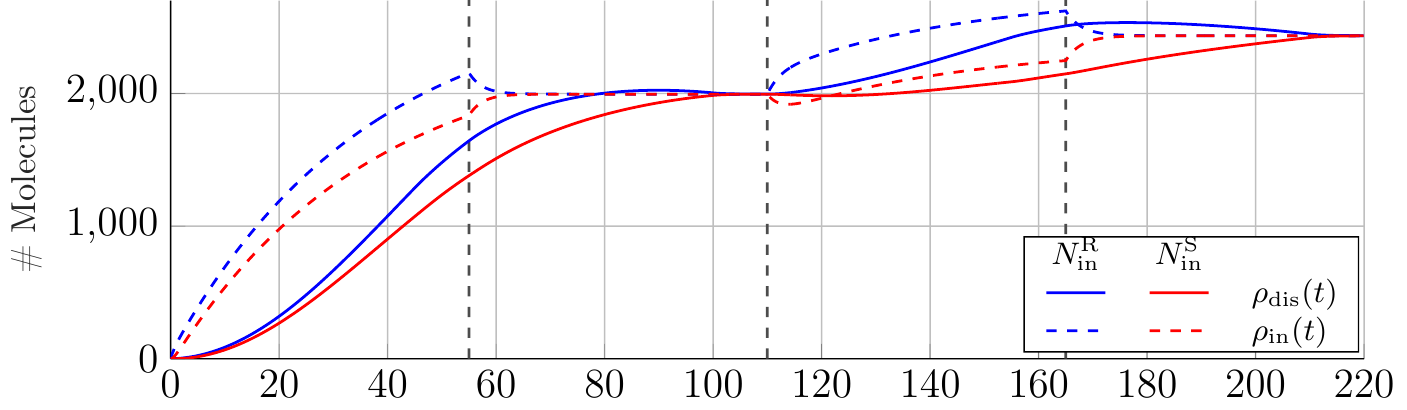} \\[-0.3em]
	\includegraphics[width=\linewidth]{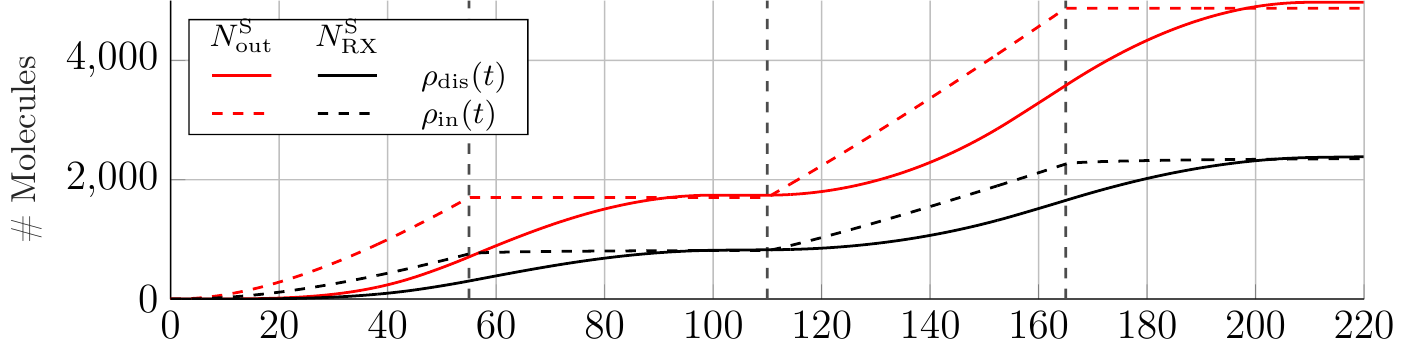}\\[-0.3em]
	\includegraphics[width=\linewidth]{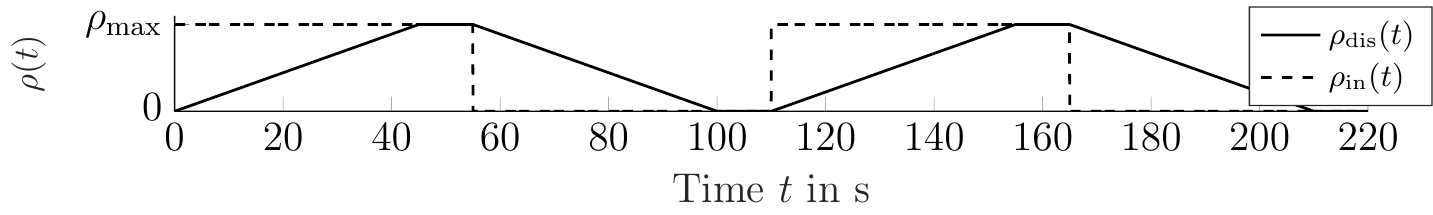}
	\vspace*{-0.7cm}
	\caption{\small Top: Number of \ac{(R)man} $\NRin$ and \ac{(S)man} molecules $\NSin$ inside the \ac{Tx}. Center: Number of molecules released by the \ac{Tx} $\NSout$ and received at the \ac{Rx} $\NSrx$. Bottom: Idealized $\rho_\mathrm{in}$ and realistic $\rho_\mathrm{dis}$ membrane switching pattern. }
	\label{fig:cr:switching}
	\vspace*{-0.7cm}
\end{figure}

Now, we investigate the impact of a non-instantaneous membrane switching pattern $\rho_\mathrm{dis}(t)$ for the practical \ac{NP} described in Section~\ref{sec:realistic}.
Therefore, we compare two switching patterns in the following. 
An idealized instantaneous switching pattern $\rho_\mathrm{in}$ with $t_\mathrm{s}\in\{0, 55, 110, 165\}\si{\second}$ and a realistic switching pattern $\rho_\mathrm{dis}$ where each membrane switch from open to close and vice versa is distributed over $t_\mathrm{dis} = 45\si{\second}$ (see bottom of Fig.~\ref{fig:cr:switching}). 
The resulting long simulation durations for realistic switching durations makes the usage of \ac{PBS} infeasible. 
Instead, we rely on the results of our proposed model as it can handle longer time sequences and we have shown in the previous sections that it matches the results from \ac{PBS} very well.
Fig.~\ref{fig:cr:switching} shows the number of \ac{(R)man} and \ac{(S)man} molecules inside the \ac{Tx} (top) and the number of molecules released from the \ac{Tx} and received at the \ac{Rx} (center). 
First, we observe that the total number of molecules inside the \ac{Tx}, released from the \ac{Tx}, and received by the \ac{Rx} increased significantly due to the long switching intervals.  
Next, we observe that the equilibrium inside the \ac{Tx} is reached faster for the instantaneous switching pattern $\rho_\mathrm{in}$ than for the practical $\rho_\mathrm{dis}$. 
Comparing the number of received molecules $\NSrx$ for $\rho_\mathrm{dis}$ (solid lines) and the instantaneous $\rho_\mathrm{in}$ (dashed lines), we observe that the individual releases of signaling molecules from the \ac{Tx} are no longer clearly distinguishable for the practical pattern $\rho_\mathrm{dis}$. 
This effect is expected, because $\rho_\mathrm{dis}$ also spreads the release of molecules from the \ac{Tx} over the switching intervals, i.e., the number of released molecules per second increases with increasing permeability. 
In contrast, for the instantaneous pattern $\rho_\mathrm{in}$ the molecule release starts and stops immediately. 
These results reveal that the non-instantaneous switching of the practical \ac{Tx} model has a significant influence on the received number of molecules. 
The resulting implication for the design of suitable detectors is left for future work.  
 \vspace*{-0.6cm}

\section{Conclusions}
\label{sec:conclusion}
\vspace*{-0.1cm}
In this paper, we proposed a novel \ac{Tx} model for \ac{MC} based on functionalized \acp{NP}, which overcomes common assumptions of existing \ac{Tx} models. 
We presented a general mathematical model and specialized the system to a realization relying on idealized assumptions and a practical realization. 
In our numerical evaluation, we investigated the signaling molecule production mechanisms and the influence of multiple, (non-)instantaneous membrane switching cycles. 
By comparison to \ac{PBS}, we confirmed the validity of \mbox{our models}.

The results of this paper are a first step towards the development of more practical \ac{Tx} models for \ac{MC}. 
Interesting topics for future work include a more detailed investigation of all design parameters for both the idealized and the practical realization of the proposed \ac{Tx} model, the evaluation of the communication theoretical measures, e.g., the bit error rate, and the development of a experimental testbed to validate the proposed models with experimental data. 
\vspace*{-0.2cm}


\bibliographystyle{ACM-Reference-Format}
\bibliography{libdat.bib}

\end{document}